%% file: GeodesicEmbedding-arXiv.tex
\documentclass[10pt,journal,compsoc]{IEEEtran}

\usepackage{ifpdf}
\ifCLASSOPTIONcompsoc
  \usepackage[nocompress]{cite}
\else
  \usepackage{cite}

\fi

\ifCLASSINFOpdf
\else
\fi

\usepackage{amsmath}
\usepackage{amssymb}
\usepackage{bm}
\usepackage{array}
\usepackage{algorithm}
\usepackage{comment}
\usepackage{algpseudocode}
\usepackage{graphicx}
\usepackage{xcolor}
\usepackage{booktabs}
\usepackage{multirow}
\usepackage{url}
\usepackage{microtype}

\newcommand\etc{etc\@ifnextchar.{}{.\@}}

\newcommand{\edit}[1]{{#1}}

\newcommand{\Rnum}[1]{\uppercase\expandafter{\romannumeral #1\relax}}

\begin{document}

\title{GeodesicEmbedding (GE):\\ A High-Dimensional Embedding Approach for Fast Geodesic Distance Queries}

\author{Qianwei~Xia,~
        Juyong~Zhang$^\dagger$,~\IEEEmembership{Member,~IEEE,}~
         Zheng~Fang,~
         Jin~Li,~
         Mingyue~Zhang,\\
         Bailin~Deng,~\IEEEmembership{Member,~IEEE,}~
         Ying~He,~\IEEEmembership{Member,~IEEE}
\IEEEcompsocitemizethanks{\IEEEcompsocthanksitem Q. Xia, J. Zhang and M. Zhang are with the School of Mathematical Sciences, University of Science and Technology of China. 
\IEEEcompsocthanksitem J. Li is with the School of Computer Science and Technology, University of Science and Technology of China.
\IEEEcompsocthanksitem Z. Fang and Y. He are with the School of Computer Science and Engineering, Nanyang Technological University.
\IEEEcompsocthanksitem B. Deng is with the School of Computer Science and Informatics, Cardiff University.}
\thanks{$^\dagger$Corresponding author. Email: {\texttt{juyong@ustc.edu.cn}}.}
}

\markboth{}
{}

\IEEEtitleabstractindextext{%
\begin{abstract}
In this paper, we develop a novel method for fast geodesic distance queries.
The key idea is to embed the mesh into a high-dimensional space, such
that the Euclidean distance in the high-dimensional space can induce the
geodesic distance in the original manifold surface. However, directly solving
the high-dimensional embedding problem is not feasible due to the large number
of variables and the fact that the embedding problem is highly nonlinear. We overcome
the challenges with two novel ideas. First, instead of taking all vertices as
variables, we embed only the saddle vertices, which greatly reduces the
problem complexity. We then compute a local embedding for each non-saddle vertex. Second, to reduce the large approximation error resulting from the purely Euclidean embedding, we propose a cascaded optimization approach that repeatedly introduces additional embedding coordinates with a non-Euclidean function to reduce the approximation residual. Using the precomputation data, our approach can determine the geodesic distance between any two vertices in near-constant time. Computational testing results show that our method is more desirable than previous geodesic distance queries methods.
\end{abstract}

}

\maketitle
\begin{figure*}[t]
            \centering
            \includegraphics[width=\textwidth]{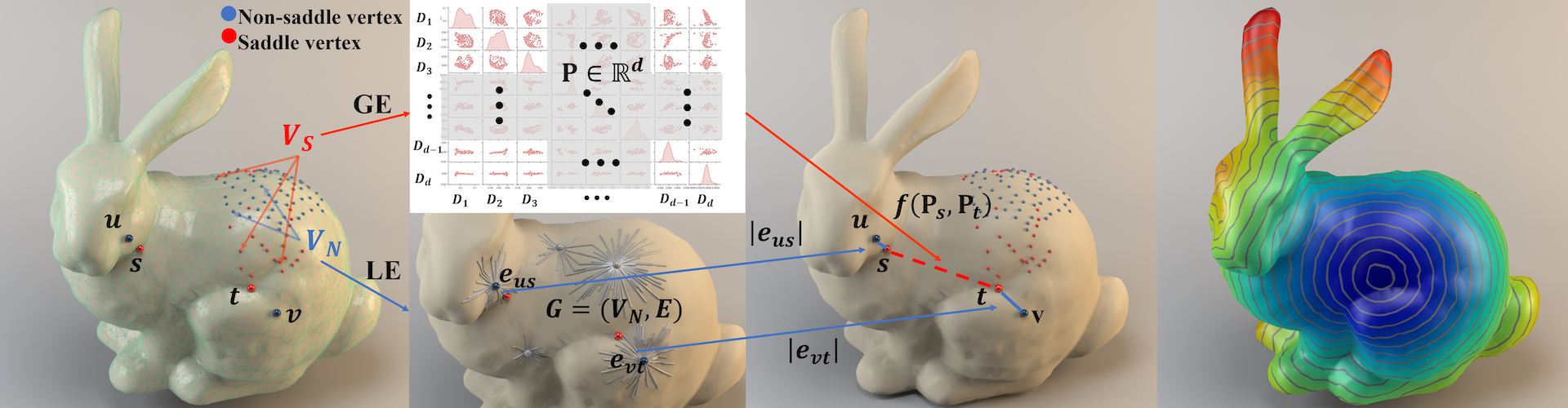}\\
            \makebox[0.245\textwidth]{(a)}
  \makebox[0.245\textwidth]{(b)}
  \makebox[0.245\textwidth]{(c)}
  \makebox[0.245\textwidth]{(d)}\\
\caption{\label{fig:teaser} Our method performs distance queries between an arbitrary pair of vertices in near constant time. (a) Distance queries for $(u,v)$. $\mathcal{V}_N$ is the set of non-saddle vertices and $\mathcal{V}_S$ is the set of saddle vertices. $u,v\in \mathcal{V}_N$ and $s,t\in \mathcal{V}_s$. (b) Top: Geodesic embedding (GE) for $\mathcal{V}_S$ and the scatter plot matrix of $\mathbf{P} \in \mathbb{R}^{d}$. $\mathbf{P}$ is the matrix for the embedded coordinates of $\mathcal{V}_S$. Bottom: Local embedding (LE) for $\mathcal{V}_N$ using Saddle Vertex Graph (SVG)~\cite{SVG}. $s$ and $t$ are the local neighbors of $u$ and $v$ respectively. The bidirectional SVG edge $e_{us},e_{vt}\in E$ stores the exact geodesic distance of $(u,s)$ and $(v,t)$. (c) The geodesic distance $d(u,v) = |e_{us}|+f(\mathbf{P}_s,\mathbf{P}_t)+|e_{vt}|$, where $\mathbf{P}_s, \mathbf{P}_t\in \mathbb{R}^{d}$ are the GE results of $s$ and $t$ from $\mathbf{P}$. (d) An illustration of a single-source-all-destinations distance field. For this bunny model ($\#V = 34835$), the average query time for a vertex pair is 0.02ms, and the relative mean error of this query is $\varepsilon = 0.7367\%$.}
	\label{Fig:Teaser}
\end{figure*}
\IEEEdisplaynontitleabstractindextext
\IEEEpeerreviewmaketitle

\input{introduction}
\input{related}
\input{algorithm}
\input{results}
\input{conclusion}

\bibliographystyle{IEEEtran}
\bibliography{GeodesicEmbedding}

\begin{IEEEbiography}[{\includegraphics[width=1in]{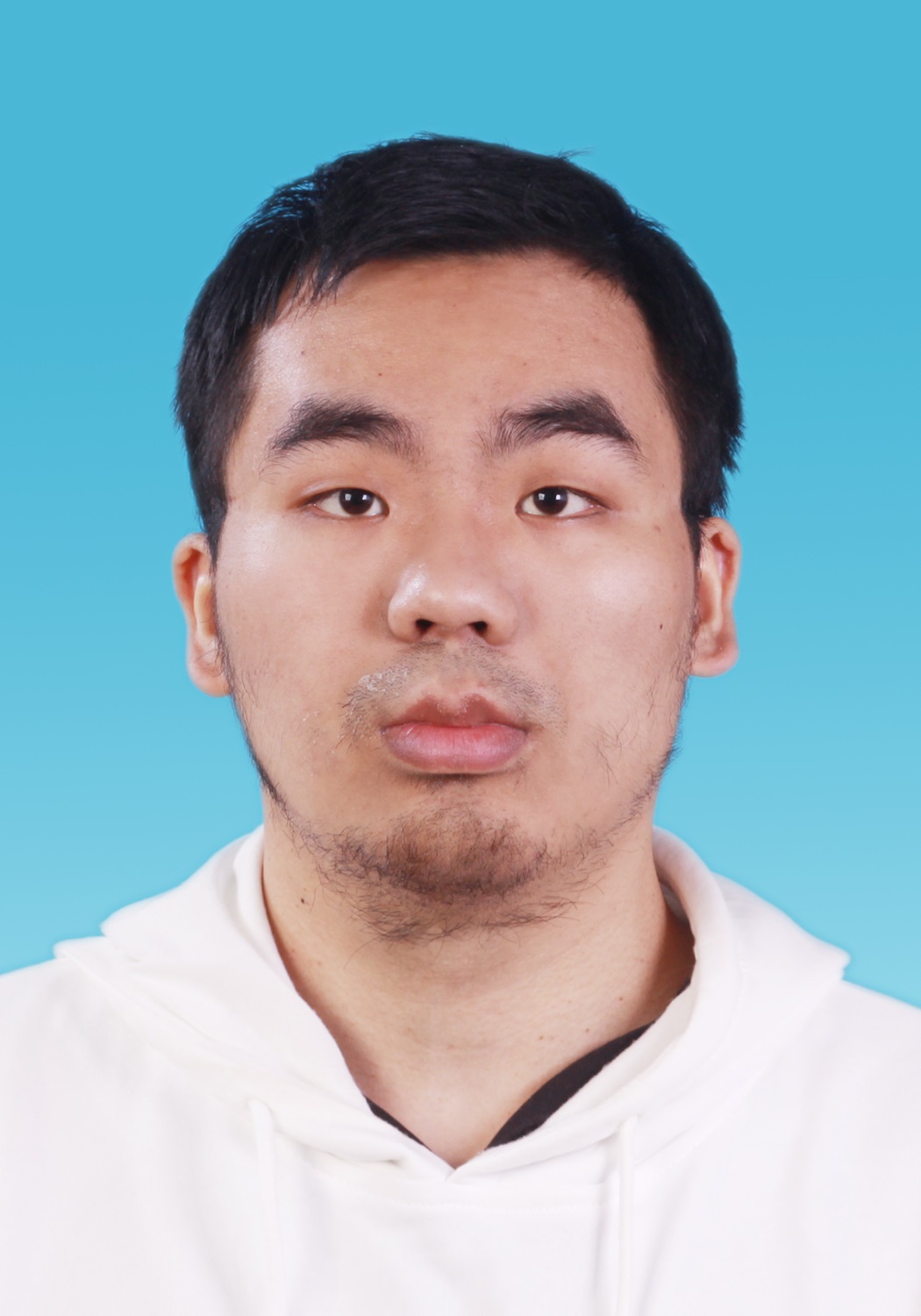}}]{Qianwei Xia} received the bachelor degree in Mathematical Sciences from University of Science and Technology of China in 2019 and  is currently working toward the master degree in Mathematical Sciences from University of Science and Technology of China. His research interests include computer graphics and numerical optimization. 
\end{IEEEbiography}

\begin{IEEEbiography}[{\includegraphics[width=1in]{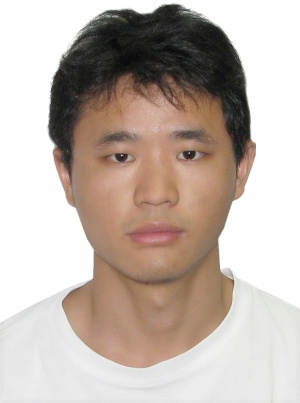}}]{Juyong Zhang}is an associate professor in the School of Mathematical Sciences at University of Science and Technology of China. He received the BS degree from the University of Science and Technology of China in 2006, and the PhD degree from Nanyang Technological University, Singapore. His research interests include computer graphics, computer vision, and numerical optimization. He is an associate editor of The Visual Computer.
\end{IEEEbiography}

\begin{IEEEbiography}[{\includegraphics[width=1in]{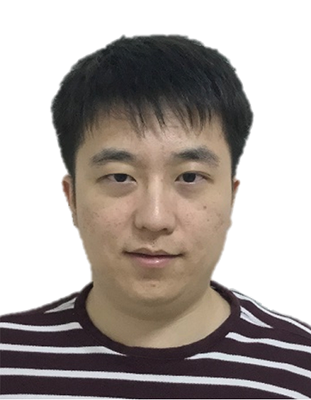}}]{Zheng Fang} received the BS degree in computer science and technology from Tianjin University, China. He is currently working toward the PhD degree with the School of Computer Science and Engineering, Nanyang Technological University, Singapore. His current research interests include computational geometry and computer graphics.\end{IEEEbiography}

\begin{IEEEbiography}[{\includegraphics[width=1in]{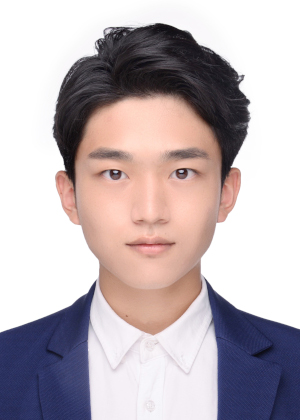}}]{Jin Li} is currently an undergraduate student at School of Computer Science and Technology at the University of Science and Technology of China. His research interests include computer graphics and computer systems.
\end{IEEEbiography}

\begin{IEEEbiography}[{\includegraphics[width=1in]{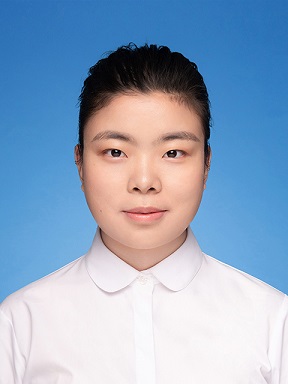}}]{Mingyue Zhang}
received the Bachelor of Science degree in Mathematical Sciences from University of Science and Technology of China in 2020 and is currently working toward the PhD degree in Mathematical Sciences from University of Science and Technology of China. Her research interests include computer vision and numerical optimization.
\end{IEEEbiography}

\begin{IEEEbiography}[{\includegraphics[width=1in]{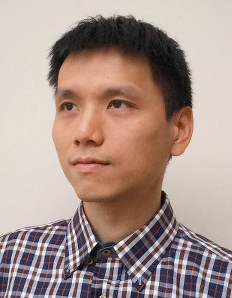}}]{Bailin Deng}
is a lecturer in the School of Computer Science and Informatics at Cardiff University. He received the BEng degree in computer software (2005) and the MSc degree in computer science (2008) from Tsinghua University (China), and the PhD degree in technical mathematics from Vienna University of Technology (Austria). His research interests include geometry processing, numerical optimization, computational design, and digital fabrication. He is a member of the IEEE.
\end{IEEEbiography}

\begin{IEEEbiography}[{\includegraphics[width=1in]{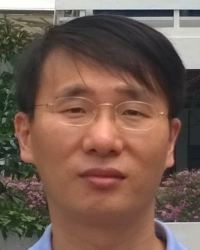}}]{Ying He} is an associate professor at School of Computer Science and Engineering, Nanyang Technological University, Singapore. He received the BS and MS degrees in electrical engineering from Tsinghua University, China, and the PhD degree in computer science from Stony Brook University, USA. His research interests fall into the general areas of visual computing and he is particularly interested in the problems which require geometric analysis and computation. For more information, visit http://www.ntu.edu.sg/home/yhe/
\end{IEEEbiography}


\vfill

\end{document}

%% file: introduction.tex
\section{Introduction}
\IEEEPARstart{T}{he} discrete geodesic problem has been investigated extensively since the seminal paper published by Mitchell et al.~\cite{Mitchell87}. Many elegant and efficient algorithms are known for computing the single-source-all-destinations (SSAD) geodesic distances in discrete domains. With the significant progress of precomputation methods, such as the heat method~\cite{CraneWW13} and the graph-based method \cite{SVG,DBLP:journals/tog/AdikusumaFH20}, SSAD can now be solved in empirically linear time.

From the perspective of applications, computing the geodesic distance between two arbitrary points, a.k.a. geodesic distance queries (GDQ), is a common scenario in shape analysis, such as  intrinsic symmetry detection~\cite{DBLP:journals/tog/XuZTLLMX09,DBLP:journals/ijcv/RavivBBK10}, surface parameterization~\cite{DBLP:journals/tvcg/ZigelmanKK02}, and shape correspondence~\cite{jain_smi06}. Some applications, such as the shape matching algorithm in~\cite{Tevs2009}, require recomputation of the geodesic distances between arbitrary points in each iteration, which would benefit from an efficient method for GDQ. Unlike SSAD, only a handful of methods have been designed for answering GDQ, such as geodesic triangle unfolding (GTU)~\cite{xin2012constant}, the Euclidean embedding metric (EEM)~\cite{PanozzoBDS13}, and the pointwise distance metric (PDM)~\cite{SolomonRGB14}. These methods, however, either have high space complexity or large error potential that limits their usage in accuracy-critical applications. For example, the naive way is to compute and store all pointwise distances. However, such an approach will consume large storage when saving the geodesic distance between all vertex pairs.

In this paper, we propose a novel method for fast geodesic distance queries. The key idea is to lift the model into a high-dimensional Euclidean space $\mathbb{R}^{d}$, so that the distance between two points in $\mathbb{R}^d$ can induce the geodesic distance on the input mesh.
The high-dimensional embedding is typically formulated as multidimensional scaling (MDS)~\cite{Mead92MDS}. However, directly solving this optimization problem is challenging due to the potentially huge number of objective function terms and the high nonlinearity of the objective function. In addition, a purely Euclidean embedding often does not approximate the pairwise geodesic distances well. We tackle these issues using two novel ideas. First, instead of lifting all vertices, we embed only the saddle vertices\footnote{A vertex is called a saddle if it has negative Gaussian curvature.} in the high-dimensional space. As pointed out in~\cite{SVG}, saddle vertices are crucial in solving the discrete geodesic problem, since they serve as the backbone of saddle vertex graphs (SVG). Based on the embedding of saddle vertices and the resulting fast evaluation of geodesic distances between them, we adapt the SVG to achieve fast GDQ for any pair of vertices.  Second, we propose a cascaded non-Euclidean embedding to gradually reduce the distance approximation error. Compared with pure Euclidean embedding, our approach can effectively improve the distance accuracy by increasing the dimensionality of the embedding vectors. An example is shown in Fig.~\ref{Fig:Teaser}. Computational results show that our method can achieve near-constant time query of geodesic distance with high accuracy only with very small storage consumption.

%% file: related.tex
\section{Related Works}

A large body of literature exists on the discrete geodesic problem. Most algorithms focus on computing SSAD on triangular meshes. Representative algorithms are wavefront propagation methods \cite{Mitchell87,Chen90,Surazhsky05,Xin09,Ying13parallel,Liu13,fwp,DBLP:journals/tog/QinHYYZ16,QH2016}, partial differential equation (PDE) methods \cite{Sethian99,Kimmel1998,DGPC,CraneWW13,DBLP:journals/cgf/BelyaevF15,PSHM}, and graph-based methods \cite{SVG,dgg,DBLP:journals/tog/AdikusumaFH20}. As mentioned above, applying these SSAD methods for GDQ is too expensive. In the following, we review only the related works on GDQ. 

The GTU method~\cite{xin2012constant} solves GDQ in constant time using preprocessed distance information. In the preprocessing step, GTU constructs the geodesic Delaunay triangulation for a fixed number of sample vertices. It computes the pairwise geodesic distance of the samples.
For each mesh vertex, the distance to the three vertices of its belonging geodesic triangle is recorded. In the query step, with the precomputed flattened distance on $\mathbb{R}^2$, GTU can answer the GDQ using an unfolding technique. GTU is conceptually simple and elegant, but it is sensitive to geometric features and only works for small-scale meshes due to its quadratic space complexity for precomputation.

Xin et al.~\cite{XIN2018Lightweight} proposed a fast query method for geodesic distance. They first construct a proximity graph using sample points from the surface. When querying the distance between two points, they augment the proximity graph with the source and target points for the query, then compute their shortest distance on the augmented graph. Our method is also based on augmenting a graph during the query, but we use an SVG for the underlying graph instead.

Panazzo et al.~\cite{PanozzoBDS13} proposed a Euclidean embedding metric that samples a representative subset $S$ of the mesh vertices and computes pairwise approximate geodesic distances for these samples. They then embed $S$ in a high-dimensional Euclidean space using metric multidimensional scaling (MMDS)~\cite{cox2000multidimensional}. After that, they embed the remaining vertices using least-square meshes~\cite{DBLP:conf/smi/SorkineC04}. EMM generates smooth distances quickly. However, their results often have large approximation errors, and their iso-distance contours are not close to geodesic distances. 

A few spectral distance techniques have been published, such as diffusion distances~\cite{Coifman7426}, commute-time distances~\cite{commutedistance} and biharmonic distances~\cite{LipmanRF10}. These distances are smooth, shape aware, and have a closed-form solution. However, they are not isotropic, so their level sets are not evenly spaced. 

Using an optimal transport map between distributions that are only nonzero at individual vertices, Solomon et al. defined a family of smooth distances on a surface transitioning from biharmonic distance to geodesic distance~\cite{SolomonRGB14}. Taking all the eigenvectors of the Laplacian matrix, the resulting distance is the exact geodesic distance. However, it is not practical to compute all the eigenvectors unless the mesh is small. Therefore, they suggest using only a few low-frequency spectral terms to approximate geodesic distances, which often yields large errors.

The idea of this high-dimension embedding is not new. Given an initial polygonal mesh, a user-specified desired number of vertices, and a parameter that specifies the desired amount of anisotropy, Levy and Bonnee~\cite{levy2013variational} generated a curvature-adapted anisotropic mesh. Their main idea is to transform the 3D anisotropic space into a higher-dimensional isotropic space, where the mesh is optimized by computing a Centroidal Voronoi Tessellation. Zhong et al.~\cite{2018Computing} presented a new method to compute a self-intersection-free, high-dimensional Euclidean embedding for surfaces and volumes equipped with an arbitrary Riemannian metric. Given an input metric, they compute a smooth, intersection-free, high-dimensional embedding of the input such that the pullback metric of the embedding matches the input metric.

%% file: algorithm.tex
\section{Proposed Model}
\label{sec:sim}
Suppose we are given a connected triangular mesh $\mathcal{M} = (\mathcal{V}, \mathcal{E}, \mathcal{F})$, where $\mathcal{V}$, $\mathcal{E}$, and $\mathcal{F}$ denote the set of vertices, edges, and faces, respectively. 
The basic idea of our approach is to compute a high-dimensional vector $\mathbf{P}_k$ for each vertex $v_k$, so that the true geodesic distance $d_{ij}$ between two vertices $v_i, v_j$ can be well approximated by a simple function $f(\mathbf{P}_i, \mathbf{P}_j)$. If the function $f$ is given, then the embedding vectors $\{\mathbf{P}_k\}$ can be determined by solving an optimization problem
\begin{equation}
\min_{\{\mathbf{P}_k\}}~~\sum_{\substack{v_i,v_j \in \mathcal{V}\\ i<j}} \Phi(f(\mathbf{P}_i, \mathbf{P}_j), d_{ij}),
\label{eq:embedding}
\end{equation}
where $\Phi(f(\mathbf{P}_i, \mathbf{P}_j), d_{ij})$ penalizes the difference between $f(\mathbf{P}_i, \mathbf{P}_j)$ and $d_{ij}$. However, performing such optimization for all vertex pairs will involve $O(|\mathcal{V}|^2)$ terms in the target function, which will require an impractical amount of storage and computational time. For better efficiency, we note that the saddle vertices can serve as relays for geodesic wavefront propagation, and a saddle vertex can be shared by multiple discrete geodesic paths connecting mesh vertices~\cite{SVG}. Therefore, we perform optimization~\eqref{eq:embedding} only on the saddle vertices to enable fast query of geodesic distance between them in $O(1)$ time. Utilizing the distance between saddle vertex pairs, we adapt the Saddle Vertex Graph (SVG) proposed in~\cite{SVG} for geodesic distance queries between an arbitrary pair of vertices with low time complexity.

In the following, we first explain our adaptation of SVG and how we use it for distance queries. Afterward, we present our method for solving the embedding problem for saddle vertices.
\begin{figure}[t!]
	\centering
	\includegraphics[width=\columnwidth]{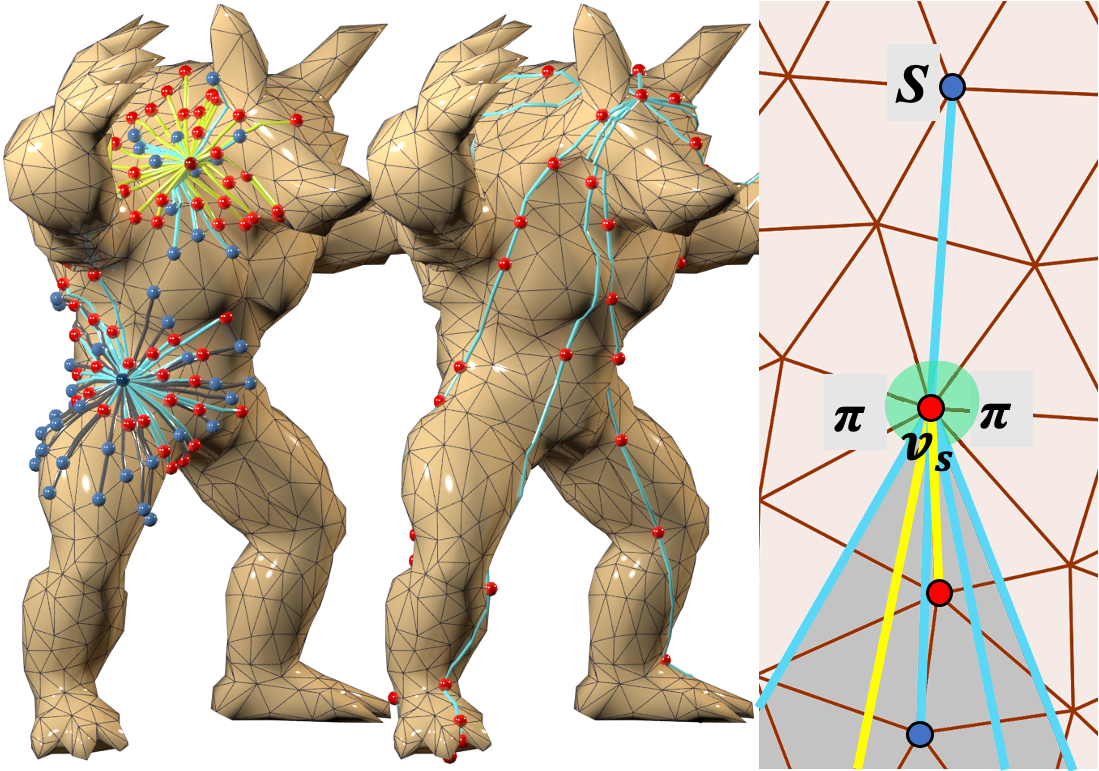}\\
	\makebox[0.333\columnwidth]{(a)}\makebox[0.333\columnwidth]{(b)}\makebox[0.333\columnwidth]{(c)}
	\caption{Illustrations of SVG construction and the properties of saddle vertex. 
		(a) Two subsets of $G$ generated from two source vertices. The red and blue spheres indicate the saddle and non-saddle vertices, respectively. The upper subset contains $E_{SS}$ in yellow and the cyan rendered $E_{NS}$. The lower subset also includes $E_{NN}$ in black. All edges are direct geodesics. 
		(b) shows that each long discrete geodesic path passes through several saddle vertices. 
		(c) Close-up view of a saddle vertex $v_s$. The geodesic paths within the shadow region of $v_s$ adopt $v_s$ as a relay vertex.
	}
	\label{fig:svg}
\end{figure}

\begin{figure*}[t]
	\centering
	\includegraphics[width=1.0\textwidth]{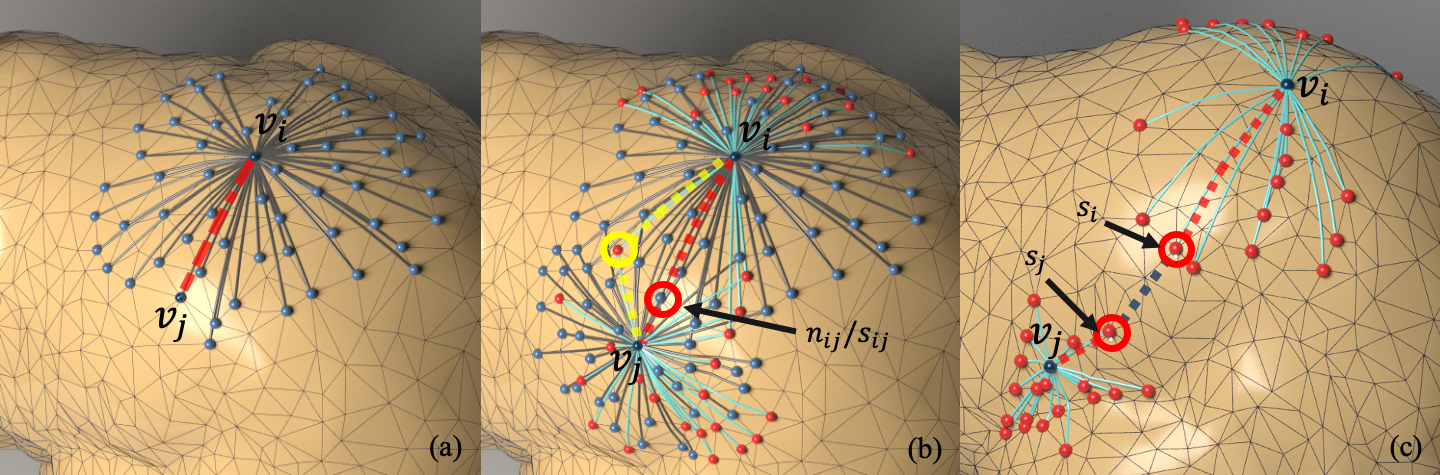}
	\caption{Solving NN-distance for $(v_i,v_j)$ using SVG. (a) The ``direct'' case. The red dotted curve indicates the shortest path, which is a direct edge in $G_{NN}$. (b) $v_i,v_j$ are ``near'' each other. $n_{ij}$ and $s_{ij}$ belong to the common set of the saddle and non-saddle neighbors of $v_i$ and $v_j$. The set intersection is computed on $G_{NN}+G_{NS}$. The yellow path illustrates the detour effect. (c) shows the ``far'' case. The black dotted line represents the embedding distance between $s_i$ and $s_j$. Only $G_{NS}$ is computed in this case.
	}
	\label{fig:NN-distance}
\end{figure*}
\subsection{SVG Construction}
The mesh vertex set $\mathcal{V}$ can be partitioned into two sets $\mathcal{V_S}$ and $\mathcal{V_N}$ containing all saddle vertices and all non-saddle vertices, respectively. Consider a geodesic wavefront propagating from a source vertex. It will split at $\mathcal{V_S}$ and continue propagating within the fan-shaped region of each saddle vertex, which is shown as the shadow region in Figure~\ref{fig:svg}. Mitchell et al.~\cite{Mitchell87} proved that a geodesic path cannot pass through a spherical vertex unless it is an endpoint or a boundary point. We define a geodesic path as \emph{direct} if it passes through no saddle vertex and \emph{indirect} otherwise. The longer a path is, the more likely it is to be indirect, but it may consist of multiple \emph{direct} paths. The saddle vertices along the discrete geodesic paths are called pseudosources in these algorithms and act as relay vertices connecting each direct geodesic path, so we can construct a graph that contains only direct paths, with saddle vertices acting as relays in the graph. The SVG proposed in~\cite{SVG} is a sparse undirected graph that utilizes saddle vertices to encode discrete geodesic paths between mesh vertices. 
The SVG for a given mesh contains all the mesh vertices as its nodes, while its edges represent direct geodesic paths connecting the nodes. Each edge has a weight that equals the geodesic distance between the two nodes it connects.
An exact SVG contains all direct geodesic paths as its edges, so that the geodesic distance between any two mesh vertices can be computed by solving a shortest path problem between their corresponding nodes on the SVG.
However, constructing an exact SVG is computationally expensive. To improve performance, Ying et al.~\cite{SVG} construct an approximate SVG by limiting the maximal degree to a parameter $K$. In their construction, the incident SVG edges for each vertex $v$ connect it to other vertices within a local geodesic disc centered at $v$ and are determined by propagation from $v$. It is shown in~\cite{SVG} that such an approximate SVG still achieves high accuracy for geodesic distance computation. The approximate SVG construction has a worst-case $O(nK^2\log K)$ and empirical $O(nK^{1.5}\log K)$ time complexity where $n=|\mathcal{V}|$ is the number of vertices, which is linear in mesh size when $K$ is fixed.

We adopt the method from~\cite{SVG} to benefit from its low complexity, with a slight modification to suit our needs. Specifically, our query method attempts to find neighboring saddle vertices for both the source and target vertices on the SVG to relay the geodesic path and to use the distance between saddle vertices where possible to benefit from its fast evaluation via embedding. Therefore, we desire a sufficient number of saddle vertex neighbors for each node in the SVG while still limiting its maximal degree.
To this end, we set another threshold parameter $K_S < K$ for the number of neighboring saddle vertices. We set $K_S = K / 3$ in our implementation.
For each vertex $v$, we terminate the the process of finding neighboring SVG nodes if we have found $K$ neighbors or $K_s$ saddle vertex neighbors. Hence the maximal degree of the resulting SVG is $O(K)$.

\subsection{Geodesic Distance Queries}
\label{sec:dis_geodesic}
The SVG constructed as mentioned above is a union of three subgraphs $G = G_{SS}\cup G_{NS}\cup G_{NN}$:
\begin{itemize}
	\item Subgraph $G_{SS} = (\mathcal{V_S},E_{SS})$ forms the skeleton of the whole SVG. Each edge of $E_{SS}$ is a direct geodesic path connecting two saddle vertices.
	\item Subgraph $G_{NS} = (\mathcal{V_S} \cup \mathcal{V_N},E_{NS})$ connects non-saddle vertices to saddle vertices.
	\item Subgraph $G_{NN} = (\mathcal{V_N},E_{NN})$ connects non-saddle vertices to non-saddle vertices.
\end{itemize}
Based on the three-tier structure of SVG, we compute the distance between any pair of vertices in different ways according to the tier it belongs to. We label the distance between two saddle vertices as the \textit{SS-distance}, the distance between two non-saddle vertices as the \textit{NN-distance}, and the distance between a non-saddle vertex and a saddle vertex as the \textit{NS-distance}.

\subsubsection{SS-distance} 
We pre-compute the embedding of $\mathcal{V_S}$ in a high-dimensional space, so that the geodesic distance between any two saddle vertices $v_i, v_j$ can be approximated using their embedded coordinates $\mathbf{P}_i, \mathbf{P}_j\in \mathbb{R}^{d}$ via a simple function $f(\mathbf{P}_i, \mathbf{P}_j)$. Afterward, the geodesic distance between two saddle vertices is evaluated by retrieving their embedded coordinates to evaluate the function $f$, which can be done in $O(1)$ total time. The definition of $f$ and the numerical method for computing the embedding will be explained in Section~\ref{eq:NumericalAlgorithm}.

\subsubsection{NN-distance.} 
If two non-saddle vertices $v_i, v_j$ are neighbors on $G_{NN}$, then their geodesic distance can be directly retrieved from their edge length. Using an associative container such as the STL map to store the neighbors for each vertex, we can determine whether $v_i$ and $v_j$ are neighbors in $O(\log K)$ time, and retrieve the distance in $O(1)$ time if this is the case. 
If $v_i$ and $v_j$ are not neighbors on the SVG, then the geodesic path between them consists of SVG edges. 
We use the following steps to determine the geodesic distance between $v_i$ and $v_j$:
\begin{itemize}
	\item When $v_i$ and $v_j$ are close, it is possible that the geodesic path is relayed by their common neighbors. Therefore, we first extract the set of all common neighbors between $v_i$ and $v_j$ on the SVG. If such a common neighbor set is not empty, then we compute the length of each path $(v_i, n_k, v_j)$ that consists of two edges connecting $v_i$ and $v_j$ to a common neighbor $n_k$, and take the shortest length of all such paths as the geodesic distance between $v_i$ and $v_j$. Since the SVG has a maximal degree $K$, this procedure can be performed in $O(K)$ time.
	\item If there is no common neighbor between $v_i$ and $v_j$, then we assume that $v_i, v_j$ are far apart enough and the geodesic path between them is relayed by saddle vertices. Thus, we first extract the set $S_i$ of saddle vertex neighbors for $v_i$, and the set $S_j$ of saddle vertex neighbors for $v_j$. If neither of them is empty, we compute for shortest distance on the graph $S_i \cup S_j \cup S_{ij}$ between $v_i$ and $v_j$ as their geodesic distance, where $S_{ij}$ is a dense graph that connects every node in $S_i$ to every node in $S_j$. We compute the shortest distance using Dijkstra's algorithm, without explicitly constructing the dense graph $S_{ij}$. Since there are $O(K)$ nodes and $O(K^2)$ edges in $S_i \cup S_j \cup S_{ij}$, and each edge length of $S_{ij}$ can be evaluated from the embedding in $O(1)$ time, the time complexity of the geodesic distance computation between $v_i$ and $v_j$ is $O(K^2)$.
	\item Finally, if at least one of $S_i$ and $S_j$ is empty, then we determine the geodesic distance between $v_i$ and $v_j$ using Dijkstra's algorithm on $G$, with time complexity $O(K n \log n)$.
\end{itemize}

\subsubsection{NS-distance.} The strategy to solve NS-distance is similar to solving \textit{NN-distance}. The difference is that the construction of local graphs only executes on the non-saddle vertex side.

\subsection{Geodesic Embedding}
\label{eq:NumericalAlgorithm}
Central to our approach is to embed the geodesic distance between saddle vertices by solving the problem~\eqref{eq:embedding}. To effectively solve this problem, we first apply high-dimensional Euclidean embedding to approximate the geodesic distance, and then apply a cascaded non-Euclidean embedding to further reduce the approximation error. 

\subsubsection{Euclidean Embedding}
We first compute an embedding vector $\mathbf{q}_k \in \mathbb{R}^m$ for each vertex $v_k$, such that the Euclidean distance $\|\mathbf{q}_i - \mathbf{q}_j  \|$ between two embedding vectors $\mathbf{q}_i, \mathbf{q}_j$ is as close as possible to the ground-truth geodesic distance $d_{ij}$ between their corresponding vertices $v_i, v_j$. To this end, we perform an optimization based on the Shape-Up formulation from~\cite{Bouaziz2012}:
\begin{equation}
\begin{aligned}
\min_{\{\mathbf{q}_k\}} \sum_{\substack{v_i,v_j \in \mathcal{V}\\ i<j}}\omega_{ij} E_{ij},
\end{aligned}
\label{Eq:Euclidean}
\end{equation}
where
\[
    E_{ij} = \min_{\|\mathbf{z}_{ij}\| = d_{ij}} \|\mathbf{q}_i - \mathbf{q}_j  - \mathbf{z}_{ij}\|^2 
\]
is the squared distance from the difference vector $\mathbf{q}_i - \mathbf{q}_j$ to the set of vectors in $\mathbf{R}^m$ with length $d_{ij}$,
and the weight $\omega_{ij} = 1 / d_{ij}^2$ is used to normalize the approximation error.
We adopt the quasi-Newton solver from~\cite{LiuBK17} for to efficiently solve~\eqref{Eq:Euclidean}.

\begin{figure}[t!]
\centering
\includegraphics[width=\columnwidth]{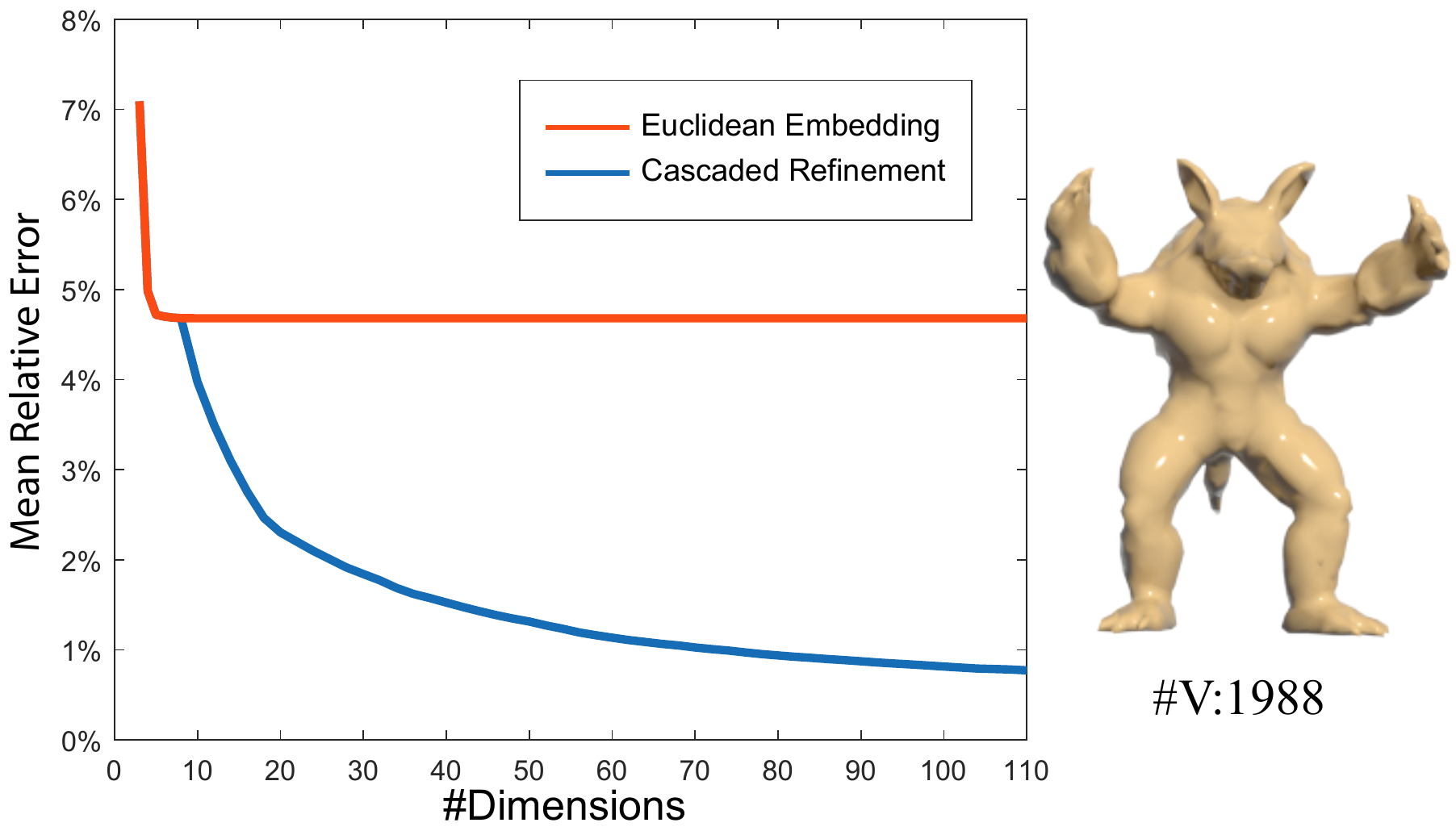}
\caption{\edit{The mean relative errors of geodesic distance (see Eq.~\eqref{eq:MeanRelError}) using Euclidean embedding, and using our approach with cascaded refinement after an initial eight-dimensional Euclidean embedding. The plots show the errors with respect to the dimensionality of the final embedding space. Our cascaded refinement can effectively reduce the approximation error with increasing dimensionality, while for pure Euclidean embedding the error stagnates after a certain threshold of dimensionality.}}
\label{fig:dimension}
\end{figure}

\subsubsection{Cascaded Non-Euclidean Embedding}
The Euclidean embedding alone may not approximate the geodesic distance very well even with a large $m$. 
Fig.~\ref{fig:dimension} shows the accuracy of Euclidean embedding on an Armadillo mesh using different values of $m$. To evaluate the accuracy, we follow~\cite{Surazhsky05} and compute the mean relative error 
\begin{equation}
	\varepsilon = \frac{1}{|\mathcal{P}|} \sum_{(v_i, v_j) \in \mathcal{P}} \varepsilon_{ij},
	\label{eq:MeanRelError}
\end{equation}
where $\mathcal{P}$ is the set of sampled vertex pairs, and $\varepsilon_{ij}$ is the relative error for the geodesic distance between two vertices $v_i$ and $v_j$:
\begin{equation}
    \varepsilon_{ij} = \frac{|d'_{ij} - d_{ij}|}{d_{ij}},
    \label{eq:RelativeError}
\end{equation}
with $d_{ij}$ and $d'_{ij}$  being the ground-truth distance and the distance computed from the embedding, respectively.
In Fig.~\ref{fig:dimension}, we use all vertex pairs on the model to evaluate the mean relative error. 
Tab.~\ref{tab:Euclidean_Embedding} also provides the values of $\varepsilon$ for different values of $m$. We can see that beyond a certain threshold, increasing the dimensionality for Euclidean embedding becomes ineffective for reducing the mean relative error. Therefore, we only perform the Euclidean embedding with a relatively small value of $m$. Afterward, we increase the dimensionality of the embedding vectors and apply a non-Euclidean function on the extra dimensions to reduce the approximation error. Our key observation is that a nonzero residual 
\[
r_1^{ij} = \|\mathbf{q}_i - \mathbf{q}_j\| - d_{ij}
\]
can be positive or negative. Therefore, we introduce a simple function for the extra embedding vector components that can attain both positive and negative values to approximate the residual; we subtract it from the existing embedding distance function to improve the approximation accuracy.
Specifically, we add two components $s_1^{(k)}, t_1^{(k)}$ to the embedding vector of $v_k$, and approximate the residual $r_1^{ij}$ for the distance between $v_i$ and $v_j$ using the following term:
\[
    (s_1^{(i)} - s_1^{(j)})^2 - (t_1^{(i)} - t_1^{(j)})^2.
\]
\begin{table}[t]
	\renewcommand{\arraystretch}{1.3}
	\caption{Mean relative error $\varepsilon$ (Eq.~\eqref{eq:MeanRelError}) of geodesic distance using Euclidean embedding (Eq.~\eqref{Eq:Euclidean}) with different dimensionality $m$ on the Armadillo model in Fig.~\ref{fig:dimension}.}
	\label{tab:Euclidean_Embedding}
	\centering
	\setlength\tabcolsep{5pt}
	\begin{tabular}{|c|c|c|c|c|c|}
		\hline
		$m$ & $\varepsilon$ & $m$ & $\varepsilon$ \\
		\hline
		3 & 5.909\% & 7 & 4.687\% \\
		\hline
		4 & 4.899\% & 8 & 4.686\% \\
		\hline
		5 & 4.739\% & 9 & 4.684\% \\
		\hline
		6 & 4.700\% & 10 & 4.680\% \\
		\hline
	\end{tabular}
\end{table}
We determine the values of $\{(s_1^{(k)}, t_1^{(k)})\}$ by solving a nonlinear least squares problem:
\begin{equation}
\min_{\{(s_1^{(k)}, t_1^{(k)})\}}   \sum_{\substack{v_i,v_j \in \mathcal{V}\\ i<j}}  \omega_{ij} \left((s_1^{(i)} - s_1^{(j)})^2 - (t_1^{(i)} - t_1^{(j)})^2 - r_1^{ij}\right)^2.
\label{eq:forres}
\end{equation}
We adopt the L-BFGS algorithm~\cite{nocedal2006numerical} in the Ceres library~\cite{ceres-solver} to solve this problem. 
Then the distance function for the embedding vectors of $v_i$ and $v_j$ becomes
\begin{equation}
    \|\mathbf{q}_i - \mathbf{q}_j\| - \left(s_1^{(i)} - s_1^{(j)}\right)^2 + \left(t_1^{(i)} - t_1^{(j)}\right)^2.
    \label{eq:NonEuclidDist}
\end{equation}
Note that both problems~\eqref{Eq:Euclidean} and \eqref{eq:forres} can be considered as instances of the following optimization problem that minimizes the distance approximation error:
\begin{equation}
    \min \sum_{\substack{v_i,v_j \in \mathcal{V}\\ i<j}}  \omega_{ij} (f(\mathbf{P}_i, \mathbf{P}_j) - d_{ij})^2
    \label{eq:ApprxErrorOptimization}
\end{equation}
where $f(\mathbf{P}_i, \mathbf{P}_j)$ denotes the function that approximates the geodesic distance between two vertices $v_i, v_j$ based on their embedding vectors $\mathbf{P}_i, \mathbf{P}_j$. 
The target function of this problem measures the overall normalized approximation error for the geodesic distance.
In problem~\eqref{Eq:Euclidean}, the vectors $\{\mathbf{P}_k\}$ are the Euclidean embedding coordinates in $\mathbf{q}_k \in \mathbf{R}^m$, and the function $f$ is the Euclidean distance. In problem~\eqref{eq:forres}, each vector $\mathbf{P}_k = (\mathbf{q}_k, s_1^{(k)}, t_1^{(k)})$ contains Euclidean and non-Euclidean components, and the function $f$ is evaluated according to Eq.~\eqref{eq:NonEuclidDist}. Moreover, as the Euclidean embedding coordinates $\{\mathbf{q}_k\}$ are the solution to problem~\eqref{Eq:Euclidean}, setting $s_1^{(k)} = t_1^{(k)} = 0$ for all vertices will result in the same overall approximation error as the pure Euclidean embedding. Therefore, the solution to problem~\eqref{eq:forres} is guaranteed to reduce the overall approximation error compared with Euclidean embedding.

\begin{table}[t]
	\renewcommand{\arraystretch}{1.3}
	\caption{
		Mean relative error $\varepsilon$ (Eq.~\eqref{eq:MeanRelError}) of geodesic distance with our cascaded refinement approach on the Armadillo model in Fig.~\ref{fig:dimension}, using an initial Euclidean embedding with dimensionality $m = 8$ and different numbers of refinement steps $l$. Our method can effectively reduce the error with an increasing $l$.}
	\label{tab:Cascaded_Refinement}
	\centering
	\setlength\tabcolsep{5pt}
	\begin{tabular}{|c|c|c|c|c|c|}
		\hline
		$l$ & $\varepsilon$ & $l$ & $\varepsilon$ & $l$ & $\varepsilon$ \\
		\hline
		0 & 4.682\% & 17 & 1.453\% & 34 & 0.9738\% \\
		\hline
		1 & 3.972\%  & 18 & 1.412\%  & 35 & 0.9575\%  \\
		\hline
		2 & 3.495\%  & 19 & 1.364\%  & 36 & 0.9441\%  \\
		\hline
		3 & 3.095\%  & 20 & 1.322\%  & 37 & 0.9288\%  \\
		\hline
		4 & 2.756\%  & 21 & 1.292\%  & 38 & 0.9159\%  \\
		\hline
		4 & 2.466\%  & 22 & 1.264\%  & 39 & 0.9024\% \\
		\hline
		6 & 2.356\%  & 23 & 1.235\% & 40 & 0.8888\%  \\
		\hline
		7 & 2.206\%  & 24 & 1.206\%  & 41 & 0.8724\%  \\
		\hline
		8 & 2.107\%  & 25 & 1.176\%  & 42 & 0.8610\%  \\
		\hline
		9 & 2.028\%  & 26 & 1.149\%  & 43 & 0.8503\%  \\
		\hline
		10 & 1.918\%  & 27 & 1.124\%  & 44 & 0.8385\%  \\
		\hline
		11 & 1.836\%  & 28 & 1.100\%  & 45 & 0.8274\%  \\
		\hline
		12 & 1.761\%  & 29 & 1.075\%  & 46 & 0.8163\%  \\
		\hline
		13 & 1.689\%  & 30 & 1.054\%  & 47 & 0.8038\%  \\
		\hline
		14 & 1.622\%  & 31 & 1.029\%  &48 & 0.7939\%  \\
		\hline
		15 & 1.563\%  & 32 & 1.011\%  & 49 & 0.7839\%  \\
		\hline
		16 & 1.504\%  & 33 & 0.9929\%  & 50 & 0.7745\%  \\
		\hline
	\end{tabular}
\end{table}
The non-Euclidean embedding optimization~\eqref{eq:forres} can be repeated to further reduce the approximation error. Concretely, using the solution to problem~\eqref{eq:forres}, we can compute a new residual for each distance $d_{ij}$ based on the distance formula~\eqref{eq:NonEuclidDist}:
\[
    r_2^{ij} = \|\mathbf{q}_i - \mathbf{q}_j\| - \left(s_1^{(i)} - s_1^{(j)}\right)^2 + \left(t_1^{(i)} - t_1^{(j)}\right)^2 - d_{ij}.
\]
Then we introduce two extra embedding vector components $s_2^{(k)}, t_2^{(k)}$ for each vertex $v_k$, and determine their values by solving an optimization problem similar to~\eqref{eq:forres}:
\begin{equation}
\min_{\{(s_2^{(k)}, t_2^{(k)})\}}  \sum_{\substack{v_i,v_j \in \mathcal{V}\\ i<j}}  \omega_{ij} \left((s_2^{(i)} - s_2^{(j)})^2 - (t_2^{(i)} - t_2^{(j)})^2 - r_2^{ij}\right)^2.
\label{eq:forres2}
\end{equation}
Then the embedding distance function becomes
\begin{align}
   & \|\mathbf{q}_i - \mathbf{q}_j\| - \left(s_1^{(i)} - s_1^{(j)}\right)^2 + \left(t_1^{(i)} - t_1^{(j)}\right)^2 \nonumber \\
   &~ - \left(s_2^{(i)} - s_2^{(j)}\right)^2 + \left(t_2^{(i)} - t_2^{(j)}\right)^2 \nonumber \\
   =~ & \|\mathbf{q}_i - \mathbf{q}_j\| - \sum_{k=1}^2 \left(s_k^{(i)} - s_k^{(j)}\right)^2
   + \sum_{k=1}^2 \left(t_k^{(i)} - t_k^{(j)}\right)^2
    \label{eq:NonEuclidDist2}
\end{align}
Following the same argument used in the previous paragraph, we can see that the introduction of the components $(s_2^{(k)}, t_2^{(k)})$ is guaranteed to reduce the overall approximation error for the geodesic distance. We repeat this process and perform $l$ rounds of non-Euclidean embedding optimization. 
The $p$-th round of optimization ($1 \leq p \leq l$) solves a problem
\begin{equation}
\min_{\{(s_p^{(k)}, t_p^{(k)})\}}  \sum_{\substack{v_i,v_j \in \mathcal{V}\\ i<j}}  \omega_{ij} \left((s_p^{(i)} - s_p^{(j)})^2 - (t_p^{(i)} - t_p^{(j)})^2 - r_p^{ij}\right)^2,
\label{eq:pthoptimization}
\end{equation}
where $r_p^{ij}$ is the residual for distance $d_{ij}$ after the previous round of optimization:
\[
    r_p^{ij} = \|\mathbf{q}_i - \mathbf{q}_j\| - \sum_{k=1}^{p-1} \left(s_k^{(i)} - s_k^{(j)}\right)^2
   + \sum_{k=1}^{p-1} \left(t_k^{(i)} - t_k^{(j)}\right)^2 - d_{ij}.
\]
After the $l$ rounds of optimization, we obtain a $(m+2l)$-dimensional embedding vector for each vertex:
\[
    \mathbf{P}_k = ( \mathbf{q}_k, \mathbf{s}_k, \mathbf{t}_k ),
\]
where $\mathbf{q}_k \in \mathbb{R}^m$ is the Euclidean components, and
\begin{align*}
\mathbf{s}_k = (s_1^{(k)}, s_2^{(k)}, \ldots, s_l^{(k)}), \mathbf{t}_k = (t_1^{(k)}, t_2^{(k)}, \ldots, t_l^{(k)}) \in \mathbb{R}^l
\end{align*}
are the non-Euclidean components. 
The embedding distance between two vectors is computed as
\begin{equation}
    f(\mathbf{P}_i, \mathbf{P}_j) = \|\mathbf{q}_i - \mathbf{q}_j\|
    - \|\mathbf{s}_i - \mathbf{s}_j\|^2 + \|\mathbf{t}_i - \mathbf{t}_j\|^2.
    \label{eq:NEEmbedding}
\end{equation}
Fig.~\ref{fig:dimension} plots the mean relative error of this embedding approach for the Armadillo model, using $m=8$ for the initial Euclidean embedding and an increase value of $l$ for the subsequent non-Euclidean refinement. Tab.~\ref{tab:Cascaded_Refinement} provides the mean relative error values for different values of $l$. We can see that the cascaded refinement effectively reduces the approximation error with increasing dimensionality of the embedding vectors.
Fig.~\ref{fig:histogram} further plots the histogram of the relative error (defined in Eq.~\eqref{eq:RelativeError}) for each type of geodesic distance (NN, NS, SS) on the Armadillo model using our method with $m = 8$ and $l = 46$.
We can see that the distance for the majority of point pairs has a relative error smaller than $2\%$, with a high concentration of errors smaller than $1\%$. This indicates a good accuracy using our method.

%% file: results.tex
\section{Results}

\begin{figure}[t!]
\centering
\includegraphics[width=\columnwidth]{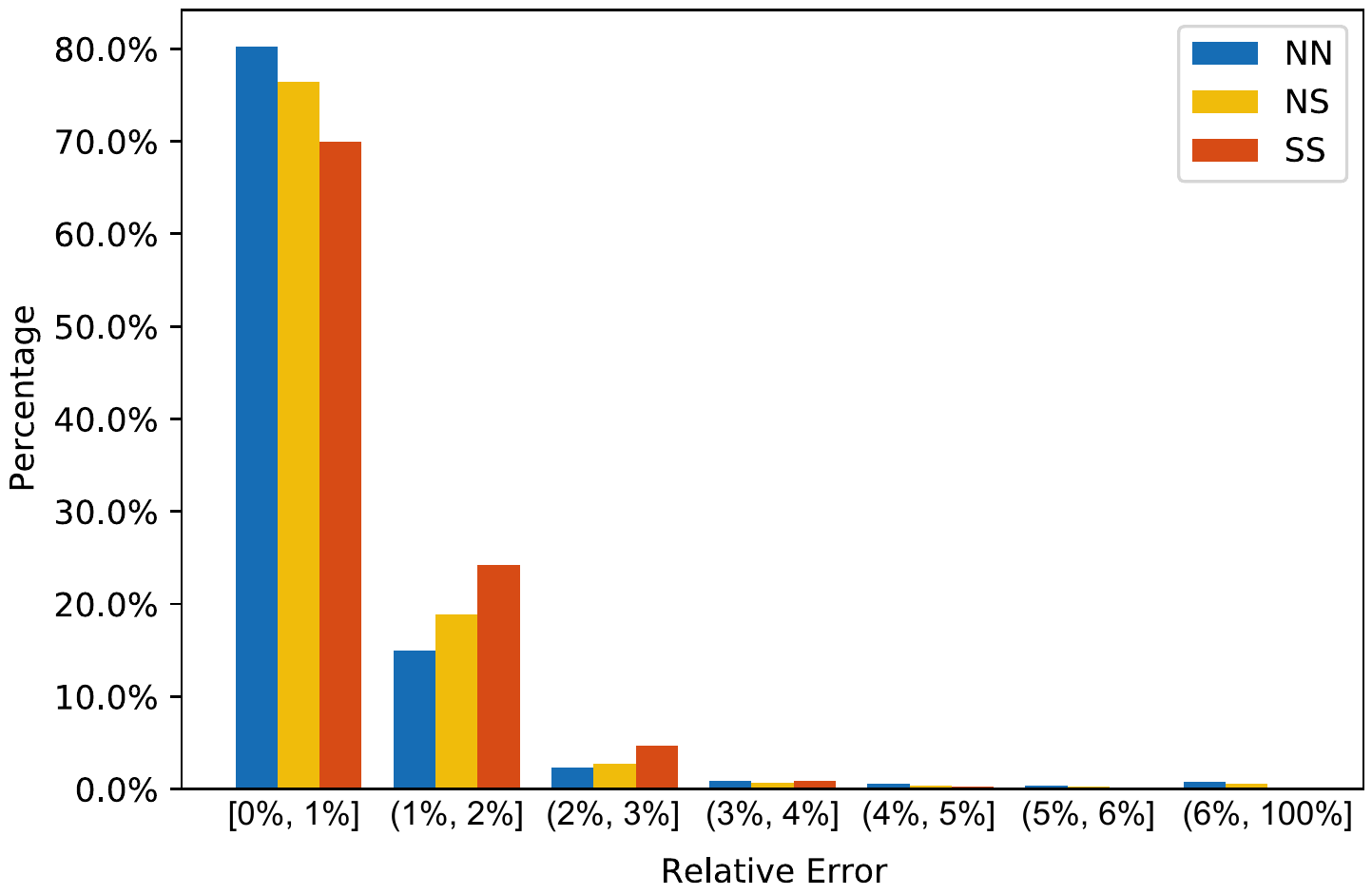}
\caption{Distribution of the relative error $\varepsilon_{ij}$ (defined in Eq.~\eqref{eq:RelativeError}) of geodesic distance for all vertex pairs on the Armadillo mesh in Fig.~\ref{fig:dimension}, using our method with $m=8$ and $l=46$.}
\label{fig:histogram}
\end{figure}

\begin{figure*}[t]
\centering
\includegraphics[width=1.0\textwidth]{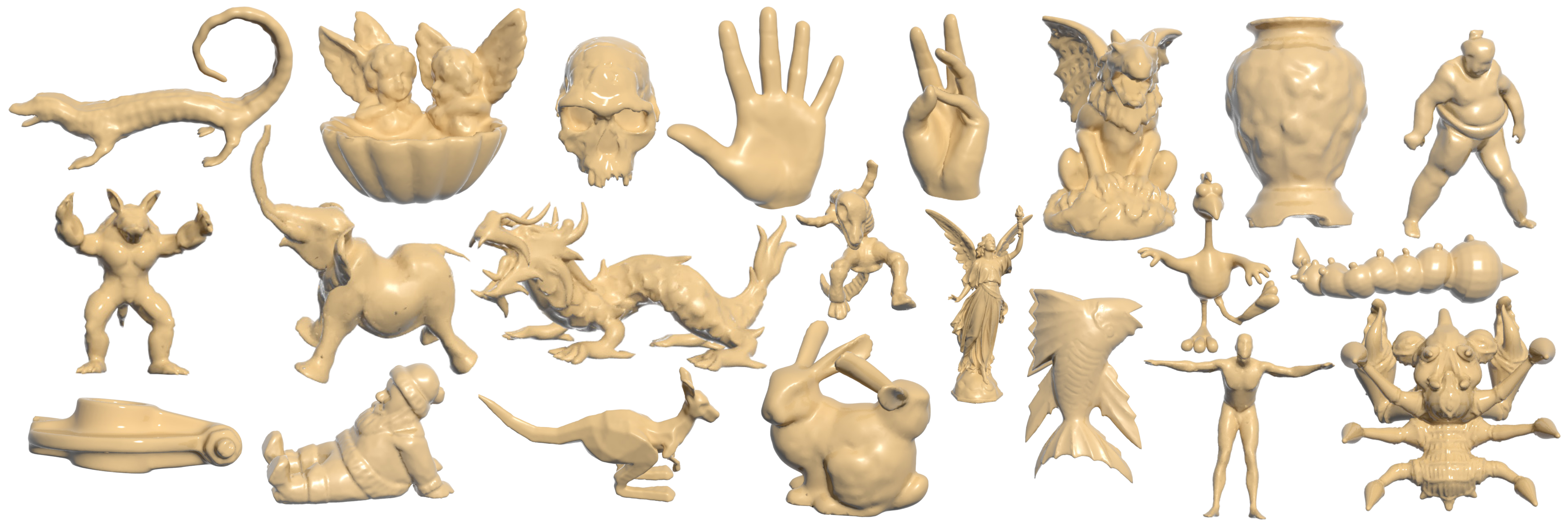}
\caption{We test and compare our method with other methods on a variety of meshes as shown in the this figure. See Table~\ref{tab:comparison} for the results.}
\label{fig:gallery}
\end{figure*}

We implemented our algorithm in C++ and used OpenMP for parallelization. All experiments were run on a workstation with an AMD Epyc 7402P at 2.8GHz and 256 GB of RAM.
To perform the embedding and evaluate the approximation error, we compute the exact geodesic distance $d_{ij}$ using the VTP method~\cite{QH2016}. Our optimization procedure requires $O(|\mathcal{V_S}|^2)$ storage space for the ground-truth geodesic distances, where $\mathcal{V_S}$ is the set of saddle vertices. After the optimization, these ground-truth distances can be discarded, and we only need to store the embeddings using $O(|\mathcal{V_S}|)$ space.

\begin{table*}[t]
    \renewcommand{\arraystretch}{1.3}
    \caption{Comparison between different methods in terms of storage, mean relative error ($\varepsilon$), and average computational time for querying geodesic distance between a random pair of vertices ($T_q$). $|\mathcal{V}|$ and $|\mathcal{V_S}|$ represent the total number of vertices and the number of saddle vertices, respectively. $T_{\textrm{SVG}}$, $T_{\textrm{VTP}}$ and $T_{\textrm{op}}$ are the time for constructing SVG, computing ground-truth distance, and performing embedding in our method, respectively}
    \label{tab:comparison}
    \centering
    \setlength\tabcolsep{1.4pt}
    \begin{small}
    \begin{tabular}{|c|c|c|c|c|c|c|c|c|c|c|c|c|c|c|c|c|c|c}
    \hline
   
    \multirow{2}{*}{
    \begin{tabular}{@{}c@{}}Model\\($|V|$, $|V_S|$)\end{tabular}}&
    \multicolumn{3}{c|}{SVG}& \multicolumn{3}{c|}{EMM-1}&
    \multicolumn{3}{c|}{EMM-2}&
    \multicolumn{6}{c|}{GE} 
    
    \\
    \cline{2-16}
    &\begin{tabular}{@{}c@{}}Storage\\(MB)\end{tabular}
    &\begin{tabular}{@{}c@{}}$T_q$\\(ms)\end{tabular}
    &$\varepsilon$
    &\begin{tabular}{@{}c@{}}Storage\\(MB)\end{tabular}
    &\begin{tabular}{@{}c@{}}$T_q$\\(ms)\end{tabular}
    &$\varepsilon$
    &\begin{tabular}{@{}c@{}}Storage\\(MB)\end{tabular}
    &\begin{tabular}{@{}c@{}}$T_q$\\(ms)\end{tabular}
    &$\varepsilon$
    &\begin{tabular}{@{}c@{}}Storage\\(MB)\end{tabular}
    &\begin{tabular}{@{}c@{}}$T_{\textrm{SVG}}$\\(s)\end{tabular}
    &\begin{tabular}{@{}c@{}}$T_q$\\(ms)\end{tabular}
    &$\varepsilon$ 
    &\begin{tabular}{@{}c@{}}$T_{\textrm{VTP}}$\\(s)\end{tabular}
    &\begin{tabular}{@{}c@{}}$T_{\textrm{op}}$\\(s)\end{tabular}\\
    \hline 
    
    \begin{tabular}{@{}c@{}}\footnotesize{Sumotori}\vspace{-0.5em}\\(9842, 5140) \end{tabular}
     & 10.80  & 1.79 & \textbf{0.354\%}
     & \textbf{0.3004} & \textbf{0.0002} & 4.723\%
     & 0.3004 & 0.0002 & 4.535\% 
     & 13.71  & 1.024 & 0.0106 & 0.7555\%   
     & 61.23 &181.2
     \\ \hline

    \begin{tabular}{@{}c@{}}\footnotesize{Kangaroo}\vspace{-0.5em}\\(9861, 5889)  \end{tabular}
     & 9.46   & 1.53 & \textbf{0.410\%}
     & \textbf{0.3009} & \textbf{0.0002} & 4.533\%
     & 0.3009 & 0.0002 & 3.897\% 
     & 12.19  & 0.837 & 0.0109 & 0.6960\%  
     &74.14  &249.4
     \\ \hline

    \begin{tabular}{@{}c@{}}\footnotesize{Click}\vspace{-0.5em}\\(9857, 5219)   \end{tabular}
     & 10.49  & 1.50 & \textbf{0.385\%}
     & \textbf{0.3008} & \textbf{0.0002} & 5.811\%
     & 0.3008 & 0.0002 & 5.057\%
     & 14.14  & 1.068 & 0.0189 & 0.6965\% 
    &63.74  &183.8
     \\ \hline

    \begin{tabular}{@{}c@{}}\footnotesize{Aligator}\vspace{-0.5em}\\(9792, 5738)    \end{tabular}
     & 9.49   & 1.57 & \textbf{0.399\%}
     & \textbf{0.2988} & \textbf{0.0002} & 3.549\%
     & 0.2988 & 0.0002 & 3.114\%
     & 12.11  & 0.937 & 0.0131 & 0.5887\%  
    &69.03 & 234.7
     \\ \hline

    \begin{tabular}{@{}c@{}}\footnotesize{Elephant}\vspace{-0.5em}\\(9979, 5401)    \end{tabular}
     & 10.37 & 1.84 & \textbf{0.537\%}
     & \textbf{0.3045} & \textbf{0.0002} & 4.820\%
     & 0.3045 & 0.0002 & 4.482\%
     & 13.15 & 1.090 & 0.0108 & 0.8204\%  
     & 63.81  &212.3
     \\ \hline

    \begin{tabular}{@{}c@{}}\footnotesize{Human}\vspace{-0.5em}\\(8559, 4757)     \end{tabular}
     & 8.75   & 1.78  & \textbf{0.465\%}
     & \textbf{0.2612} & \textbf{0.0002} & 4.798\%
     & 0.2612 & 0.0002 & 4.270\%
     & 11.24  & 0.922 & 0.0190 & 0.5905\%  
     &47.03 & 163.8
     \\ \hline

    \begin{tabular}{@{}c@{}}\footnotesize{Santa}\vspace{-0.5em}\\(20000, 11015)    \end{tabular}
     & 21.28  & 4.72 & \textbf{0.275\%}
     & \textbf{0.6104} & \textbf{0.0001} & 5.473\%
     & 0.6104 & 0.0003 & 5.179\%
     & 26.85  & 1.767 & 0.0234 & 0.7557\%  
    & 313.27  &858.4
     \\ \hline

    \begin{tabular}{@{}c@{}}\footnotesize{Hand1}\vspace{-0.5em}\\(10022, 5601)    \end{tabular}
     & 10.39  & 2.14 & \textbf{0.319\%}
     & \textbf{0.3058} & \textbf{0.0002} & 6.682\%
     & 0.3058 & 0.0002 & 5.798\%
     & 14.32 & 1.032 & 0.0179 & 0.7565\%  
     &82.48 & 214.1
     \\ \hline

    \begin{tabular}{@{}c@{}}\footnotesize{Hand2}\vspace{-0.5em}\\(10022, 5636)     \end{tabular}
     & 10.42  & 2.12  & \textbf{0.295\%}
     & \textbf{0.3058} & \textbf{0.0002} & 5.627\%
     & 0.3058 & 0.0002 & 5.188\%
     & 14.71  & 0.921 & 0.0177 & 0.7725\% 
     &91.69  &214.2
     \\ \hline

    \begin{tabular}{@{}c@{}}\footnotesize{Weedle}\vspace{-0.5em}\\(9927, 4495)     \end{tabular}
     &12.73 &2.18 & \textbf{0.283\%}
     & \textbf{0.3029} & \textbf{0.0002} &4.55\% 
     &0.3029&0.0002&4.52\%
     &14.96 & 1.054 & 0.0214 &0.667\%  
     &54.37  &140.6
     \\ \hline
     
     \begin{tabular}{@{}c@{}}\footnotesize{Gargoyle}\vspace{-0.5em}\\(6002, 3264)   \end{tabular}
     & 5.91  &1.30 & \textbf{0.434\%}
     & \textbf{0.1832}& \textbf{0.0001} &5.604\% 
     & 0.1832&0.0001&5.287\%
     & 7.365 & 0.595 & 0.0124 &1.098\%  
    &18.18 & 69.5
     \\ \hline  
     
     \begin{tabular}{@{}c@{}}\footnotesize{Dragon}\vspace{-0.5em}\\(5250, 2841)   \end{tabular}
     &4.90  &0.82 & \textbf{0.551\%}
     & \textbf{0.1602} & \textbf{0.0001} &3.976\% 
     &0.1602&0.0001&3.737\%
     &5.964 & 0.538 & 0.0162 &0.9902\%  
    &13.56  &56.9
     \\ \hline  
     
     \begin{tabular}{@{}c@{}}\footnotesize{Homohabilis}\vspace{-0.5em}\\(5002, 2734)   \end{tabular}
     &4.93  &0.95 & \textbf{0.428\%}
     & \textbf{0.1526} & \textbf{0.0001} &6.382\% 
     &0.1526&0.0001&6.224\%
     &6.133 & 0.522 & 0.0158 &0.8201\%  
     &13.19  &50.8
     \\ \hline  
     
     \begin{tabular}{@{}c@{}}\footnotesize{Vase}\vspace{-0.5em}\\(10002, 5857)   \end{tabular}
     &10.04 &2.04 & \textbf{0.395\%}
     & \textbf{0.3052} & \textbf{0.0002} &6.115\% 
     &0.3052&0.0002&5.84\%
     &12.76 & 0.949 & 0.0105 &0.6767\%  
     &84.17  &213.6
     \\ \hline  
     
     \begin{tabular}{@{}c@{}}\footnotesize{Angels}\vspace{-0.5em}\\(7527, 3787)   \end{tabular}
     &8.47 &1.89 & \textbf{0.316\%}
     & \textbf{0.2297} & \textbf{0.0001} &5.04\% 
     &0.2297&0.0002&4.786\%
     &10.34 & 0.800 & 0.0123 &0.8233\%  
     &31.35 & 100.1
     \\ \hline  
     
     \begin{tabular}{@{}c@{}}\footnotesize{Rockerarm}\vspace{-0.5em}\\(9423, 4340)   \end{tabular}
     &12.06 &1.42 & \textbf{0.357\%}
     &\textbf{0.2876} & 0.0002&7.162\% 
     &0.2876&\textbf{0.0001}&6.847\%
     &18.89 & 1.086 &  0.0185 &0.8233\%  
     & 76.67  &149.7
     \\ \hline  
     
     \begin{tabular}{@{}c@{}}\footnotesize{TwoHeadedBunny}\vspace{-0.5em}\\(18111, 9503)   \end{tabular}
     &10.74 &1.92 &\textbf{0.571\%}
     &\textbf{0.2924}& \textbf{0.0001}&5.759\% 
     &0.2924&0.0002&5.576\%
     &14.71 & 1.813 & 0.0228 &1.091\%  
     &283.51  &667.0
     \\ \hline  
     
     \begin{tabular}{@{}c@{}}\footnotesize{Thundercrab}\vspace{-0.5em}\\(9584, 4975)   \end{tabular}
    &19.63 &3.54 & \textbf{0.429\%}
    &\textbf{0.5527} & \textbf{0.0001} &4.804\% 
    &0.5527&0.0002&4.586\%
    &24.54 & 1.007 & 0.0176 &0.8927\%  
    &47.93  &181.3
    \\ \hline 
     
     \begin{tabular}{@{}c@{}}\footnotesize{Armadillo}\vspace{-0.5em}\\(1988, 973)   \end{tabular}
    &2.090 &0.50 & \textbf{0.346\%}
    &\textbf{0.0607} & \textbf{0.0001} &4.602\% 
    &0.0607&0.0001&4.525\%
    &2.441 & 0.259 & 0.0194 &0.7106\%  
   &1.75  &2.9
    \\ \hline
      
    \begin{tabular}{@{}c@{}}\footnotesize{Lucy}\vspace{-0.5em}\\(11906, 6146)   \end{tabular}
    &12.64 &2.23 & \textbf{0.420\%}
    &\textbf{0.3633} & \textbf{0.0001}&4.868\% 
    &0.3633&0.0002&4.57\%
    &15.38 &1.294  & 0.0140&0.823\%  
    & 83.08  &272.0
    \\ \hline
    \end{tabular}
    \end{small}
\end{table*}

In all our experiments, we set the dimensionality of the initial Euclidean embedding to $m=8$ because a larger $m$ provides little reduction of the approximation error. This is also the recommended setting of the method in~\cite{PanozzoBDS13}.
For the cascaded refinement, more rounds of optimization~\eqref{eq:pthoptimization} can reduce the approximation error but also increase the computational time. We set $l = 46$ in our implementation to achieve a balance between speed and accuracy.
The parameter $K_S$ also influences the approximation accuracy, query time, and storage. In general, a larger $K_S$ results in better accuracy because it increases the number of candidate paths between two non-saddle vertices via saddle vertices, which helps to improve the accuracy of NN-distances. Unfortunately, a larger $K_S$ leads to longer computational time and higher storage requirements. We set $K_S = 20$ in all experiments.

In the following, we compare the GDQ performance between our method and state-of-the-art methods, including the SVG method~\cite{SVG} (with $K = 40$), and the method from~\cite{PanozzoBDS13}, which we denote as EMM. EMM first samples a user-specified number of points from the mesh model, then applies metric multidimensional scaling (MMDS)~\cite{cox2000multidimensional} to perform embedding in $\mathbb{R}^{d}$. To complete the embedding for the remaining vertices, EMM interpolates a smooth mesh in the $d$-dimensional space by solving a Poisson equation. Following that paper, we set the number of sample points to 1000 and the embedding dimension to $d = 8$ by default.

We compare the methods using the models shown in Fig.~\ref{fig:gallery}, and the results are shown in Tab.~\ref{tab:comparison}.
We evaluate each method's mean relative error (as defined in Eq.~\eqref{eq:MeanRelError}), average distance queries time for a pair of vertices, and storage consumption.
For our method, we also provide the computational time for the ground-truth distance and for the embedding.
For each model, we evaluate the mean relative error and the average distance queries time on the same set of $10K$ vertex pairs, and we use the VTP method~\cite{QH2016} to compute the ground-truth geodesic distance for evaluating the mean relative error.

Tab.~\ref{tab:comparison} shows that our method can evaluate the geodesic distance between an arbitrary pair of vertices in about 0.02ms, with low storage consumption. Compared with our method, the SVG method requires less storage space for the graph information of GDQ and can achieve better accuracy, but its average distance queries time is about two orders of magnitude longer than our method. We test two versions of the EMM method: EMM-1, which samples 500 vertices, and EMM-2, which samples the same number of vertices ($|\mathcal{V_S}|$) as our method, with the embedding dimensionality set to 8 in both cases. We can see that EMM-2 can achieve slightly better results than EMM-1. The mean relative errors of the EMM methods are about $5\%$, while the mean relative error of our method is less than $1\%$. 

\begin{figure*}[t]
\centering
\includegraphics[width=1.0\textwidth]{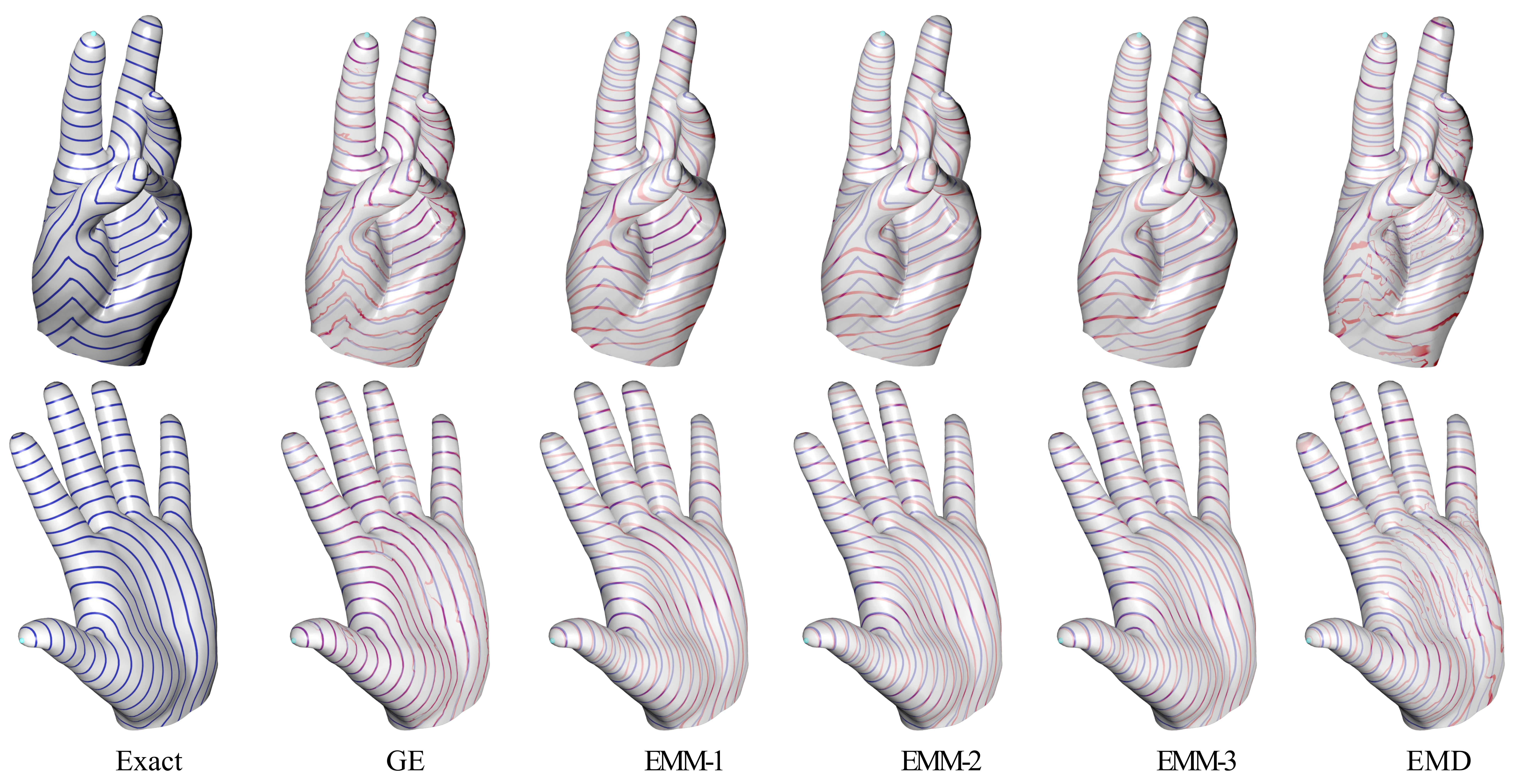}
\caption{Comparison between the iso-contours of geodesic distance from the same vertex, computed using different methods. The first column shows the ground-truth iso-contours computed using VTP~\cite{QH2016}. To visually compare the quality, we overlay the iso-contours from each method with the ground truth. The  mean relative errors are: (top row) GE $0.9336\%$, EMM-1 $5.8782\%$, EMM-2 $4.990\%$, EMM-3 $5.0134\%$, and EMD $26.7218\%$; (bottom row) GE $0.6414\%$, EMM-1 $3.7507\%$, EMM-2 $3.8316\%$, EMM-3 $3.8322\%$, and EMD $11.4222\%$.
}
\label{fig:errormap}
\end{figure*}

In Fig.~\ref{fig:errormap}, we also show the iso-contours of exact geodesic distance and the distance functions computed by different methods. Besides the two variants EMM-1 and EMM-2 mentioned above for the EMM method, we also include another variant EMM-3, with the number of sample points being the same as the number of saddle points used in our method and the embedding dimension increased to 100.
Also included in Fig.~\ref{fig:errormap} for comparison is the method from~\cite{SolomonRGB14} based on the earth mover's distance (EMD). 
We compute the distance using the code provided by the authors of~\cite{SolomonRGB14} and use a basis size 100.
Figure~\ref{fig:errormap} shows that our results have low approximation errors and our iso-contours largely follow those of the exact contours, but our distance functions are less smooth than those from the other methods. The other methods achieve smoother iso-contours, but at the cost of higher approximation errors. Therefore, our method complements existing approaches and can be desirable in applications that require higher accuracy.

\begin{table}[t]
    \renewcommand{\arraystretch}{1.3}
    \caption{Comparison between representative methods in terms of accuracy, need for precomputation (PC), space complexity (SC), and time complexities (TC) for SSAD and GDQ. Here $K$ is the maximal degree of SVG; $m$ is the number of samples on the input mesh of GTU; $\varepsilon'$ is the accuracy parameter of DGG; 
    Acronyms are PC (precomputation), SC (space complexity) and TC (time complexity).  }
    \label{tab:repmethods}
    \centering
    \setlength\tabcolsep{2pt}
    \begin{footnotesize}
    \begin{tabular}{|c|c|c|c|c|c|}
    \hline
        Method & Accuracy & PC & SC & TC (SSAD) & TC (GDQ) \\
        \hline
        MMP~\cite{Mitchell87} & Exact & No & $O(n^2)$ & $O(n^2\log n)$ &  $O(n^2\log n)$\\
        CH~\cite{Chen90} & Exact & No & $O(n)$ & $O(n^2)$ & $O(n^2)$ \\
        ICH~\cite{Xin09} & Exact & No & $O(n)$ & $O(n^2\log n)$ & $O(n^2\log n)$\\
        FWP~\cite{fwp} & Exact & No & $O(n)$ & $O(n^2)$ & $O(n^2)$\\
        VTP~\cite{QH2016} & Exact & No & $O(n)$ & $O(n^2)$& $O(n^2)$\\
        FMM~\cite{Kimmel1998} & Approx & No & $O(n)$ & $O(n\log n)$& $O(n\log n)$ \\
        DGPC~\cite{DGPC} & Approx & No & $O(n)$ & $O(n\log n)$ &  $O(n\log n)$ \\
        HM~\cite{CraneWW13} & Approx & Yes  & $O(n)$ & $O(n)$ & $O(n)$\\
        PSHM~\cite{PSHM} & Approx & No & $O(n)$ & $O(n)$ & $O(n)$ \\
        SVG~\cite{SVG} & Approx & Yes &  $O(Kn)$ & $O(Kn\log n)$ & $O(Kn\log n)$\\
        DGG~\cite{DBLP:journals/tog/AdikusumaFH20} & Approx & Yes & $O(\frac{n}{\sqrt{\varepsilon'}})$ & $O(n)$ & $O(n)$ \\
        GTU~\cite{xin2012constant} & Approx & Yes & $O(m^2+n)$ &
        $O(n)$ &  $O(1)$\\
        GE (Ours) & Approx & Yes & $O(n)$ & $O(n)$ & $O(1)$  \\
        \hline
    \end{tabular}
    \end{footnotesize}
\end{table}

%% file: conclusion.tex
\section{Conclusions \& Future Work}
We developed a new method to embed triangular meshes in a high-dimensional space so that the Euclidean distances can approximate the geodesic distances well. Our method is based on two novel ideas. 
First, instead of taking all vertices as variables, we embed only the saddle vertices, which greatly reduces the problem complexity. We then compute a local embedding for each non-saddle vertex. 
Second, to solve the large residual issue, we propose a cascaded optimization method that can effectively reduce the residual in a step-by-step manner. For both the global and local embeddings, we can quickly compute the geodesic distance between any two vertices in near-constant time. Our computational results show that this new method is more accurate than previous geodesic distance queries methods. Tab.~\ref{tab:repmethods} compares the characteristics of our method with representative methods for SSAD and GDQ.

Our method also has limitations. First, although the distance computed by our method is quite close to the exact geodesic distance, its level sets are not smooth, and there is no guarantee that it is a metric. How to add these constraints into the GE framework is an interesting research topic. Second, the numerical optimization algorithm currently used to solve this model is still computationally expensive for large meshes.  Third, we use saddle vertices as embedding elements. However, some meshes may have very few saddle vertices, yielding a poor result. Discrete geodesic graph \cite{dgg} (DGG), which uses any kind of vertex as relays to approximate long geodesic paths, may be used instead. A potential future project is to develop a lightweight numerical solver with improved efficiency for embedding.

\section*{Acknowledgments}
This work was supported by National Natural Science Foundation of China (No. 62122071), the Youth Innovation Promotion Association CAS (No. 2018495), ``the Fundamental Research Funds for
the Central Universities'' (No. WK3470000021), Guangdong International Science and Technology Cooperation Project (No. 2021A0505030009), and Singapore MOE RG26/17.

%% file: GeodesicEmbedding-arXiv.bbl
\begin{thebibliography}{10}
\providecommand{\url}[1]{#1}
\csname url@samestyle\endcsname
\providecommand{\newblock}{\relax}
\providecommand{\bibinfo}[2]{#2}
\providecommand{\BIBentrySTDinterwordspacing}{\spaceskip=0pt\relax}
\providecommand{\BIBentryALTinterwordstretchfactor}{4}
\providecommand{\BIBentryALTinterwordspacing}{\spaceskip=\fontdimen2\font plus
\BIBentryALTinterwordstretchfactor\fontdimen3\font minus
  \fontdimen4\font\relax}
\providecommand{\BIBforeignlanguage}[2]{{%
\expandafter\ifx\csname l@#1\endcsname\relax
\typeout{** WARNING: IEEEtran.bst: No hyphenation pattern has been}%
\typeout{** loaded for the language `#1'. Using the pattern for}%
\typeout{** the default language instead.}%
\else
\language=\csname l@#1\endcsname
\fi
#2}}
\providecommand{\BIBdecl}{\relax}
\BIBdecl

\bibitem{SVG}
X.~Ying, X.~Wang, and Y.~He, ``Saddle vertex graph ({SVG}): A novel solution to
  the discrete geodesic problem,'' \emph{ACM Transactions on Graphics},
  vol.~32, no.~6, pp. 170:1--12, 2013.

\bibitem{Mitchell87}
J.~S. Mitchell, D.~M. Mount, and C.~H. Papadimitriou, ``The discrete geodesic
  problem,'' \emph{SIAM Journal on Computing}, vol.~16, no.~4, pp. 647--668,
  1987.

\bibitem{CraneWW13}
K.~Crane, C.~Weischedel, and M.~Wardetzky, ``Geodesics in heat: {A} new
  approach to computing distance based on heat flow,'' \emph{{ACM} Trans.
  Graph.}, vol.~32, no.~5, pp. 152:1--152:11, 2013.

\bibitem{DBLP:journals/tog/AdikusumaFH20}
Y.~Y. Adikusuma, Z.~Fang, and Y.~He, ``Fast construction of discrete geodesic
  graphs,'' \emph{{ACM} Trans. Graph.}, vol.~39, no.~2, pp. 14:1--14:14, 2020.

\bibitem{DBLP:journals/tog/XuZTLLMX09}
K.~Xu, H.~Zhang, A.~Tagliasacchi, L.~Liu, G.~Li, M.~Meng, and Y.~Xiong,
  ``Partial intrinsic reflectional symmetry of 3d shapes,'' \emph{{ACM} Trans.
  Graph.}, vol.~28, no.~5, p. 138, 2009.

\bibitem{DBLP:journals/ijcv/RavivBBK10}
D.~Raviv, A.~M. Bronstein, M.~M. Bronstein, and R.~Kimmel, ``Full and partial
  symmetries of non-rigid shapes,'' \emph{Int. J. Comput. Vis.}, vol.~89,
  no.~1, pp. 18--39, 2010.

\bibitem{DBLP:journals/tvcg/ZigelmanKK02}
G.~Zigelman, R.~Kimmel, and N.~Kiryati, ``Texture mapping using surface
  flattening via multidimensional scaling,'' \emph{{IEEE} Trans. Vis. Comput.
  Graph.}, vol.~8, no.~2, pp. 198--207, 2002.

\bibitem{jain_smi06}
A.~Clements and H.~Zhang, ``Robust 3d shape correspondence in the spectral
  domain,'' in \emph{Proc. of Shape Modeling International}, 2006, pp.
  118--129.

\bibitem{Tevs2009}
A.~Tevs, M.~Bokeloh, M.~Wand, A.~Schilling, and H.-P. Seidel, ``Isometric
  registration of ambiguous and partial data,'' in \emph{2009 IEEE Conference
  on Computer Vision and Pattern Recognition}, 2009, pp. 1185--1192.

\bibitem{xin2012constant}
S.-Q. Xin, X.~Ying, and Y.~He, ``Constant-time all-pairs geodesic distance
  query on triangle meshes,'' in \emph{Proceedings of the ACM SIGGRAPH
  Symposium on Interactive 3D Graphics and Games}, 2012, pp. 31--38.

\bibitem{PanozzoBDS13}
D.~Panozzo, I.~Baran, O.~Diamanti, and O.~Sorkine{-}Hornung, ``Weighted
  averages on surfaces,'' \emph{{ACM} Trans. Graph.}, vol.~32, no.~4, pp.
  60:1--60:12, 2013.

\bibitem{SolomonRGB14}
J.~Solomon, R.~M. Rustamov, L.~J. Guibas, and A.~Butscher, ``Earth mover's
  distances on discrete surfaces,'' \emph{{ACM} Trans. Graph.}, vol.~33, no.~4,
  pp. 67:1--67:12, 2014.

\bibitem{Mead92MDS}
A.~Mead, ``Review of the development of multidimensional scaling methods,''
  \emph{Journal of the Royal Statistical Society: Series D (The Statistician)},
  vol.~41, no.~1, pp. 27--39, 1992.

\bibitem{Chen90}
J.~Chen and Y.~Han, ``Shortest paths on a polyhedron,'' in \emph{Proceedings of
  the Sixth Annual Symposium on Computational Geometry}, ser. SCG '90.\hskip
  1em plus 0.5em minus 0.4em\relax New York, NY, USA: ACM, 1990, pp. 360--369.

\bibitem{Surazhsky05}
V.~Surazhsky, T.~Surazhsky, D.~Kirsanov, S.~J. Gortler, and H.~Hoppe, ``Fast
  exact and approximate geodesics on meshes,'' \emph{ACM Trans. Graph.},
  vol.~24, no.~3, pp. 553--560, 2005.

\bibitem{Xin09}
S.-Q. Xin and G.-J. Wang, ``Improving {C}hen and {H}an's algorithm on the
  discrete geodesic problem,'' \emph{ACM Trans. Graph.}, vol.~28, no.~4, pp.
  104:1--104:8, 2009.

\bibitem{Ying13parallel}
X.~Ying, S.-Q. Xin, and Y.~He, ``Parallel {C}hen-{H}an ({PCH}) algorithm for
  discrete geodesics,'' \emph{ACM Transactions on Graphics}, vol.~33, no.~1,
  pp. 9:1--9:11, 2014.

\bibitem{Liu13}
Y.-J. Liu, ``Exact geodesic metric in 2-manifold triangle meshes using
  edge-based data structures,'' \emph{Computer-Aided Design}, vol.~45, no.~3,
  pp. 695--704, 2013.

\bibitem{fwp}
C.-X. Xu, T.~Y. Wang, Y.-J. Liu, L.~Liu, and Y.~He, ``Fast wavefront
  propagation ({FWP}) for computing exact geodesic distances on meshes,''
  \emph{IEEE Transactions on Visualization and Computer Graphics}, vol.~21,
  no.~7, pp. 822--834, 2015.

\bibitem{DBLP:journals/tog/QinHYYZ16}
Y.~Qin, X.~Han, H.~Yu, Y.~Yu, and J.~Zhang, ``Fast and exact discrete geodesic
  computation based on triangle-oriented wavefront propagation,'' \emph{{ACM}
  Trans. Graph.}, vol.~35, no.~4, pp. 125:1--125:13, 2016.

\bibitem{QH2016}
------, ``Fast and exact discrete geodesic computation based on
  triangle-oriented wavefront propagation,'' \emph{ACM Trans. Graph.}, vol.~35,
  no.~4, pp. 125:1--125:13, 2016.

\bibitem{Sethian99}
J.~A. Sethian, ``Fast marching methods,'' \emph{{SIAM} Review}, vol.~41, no.~2,
  pp. 199--235, 1999.

\bibitem{Kimmel1998}
R.~Kimmel and J.~A. Sethian, ``Computing geodesic paths on manifolds,''
  \emph{Proceedings of the National Academy of Sciences}, vol.~95, no.~15, pp.
  8431--8435, 1998.

\bibitem{DGPC}
E.~L. Melvær and M.~Reimers, ``Geodesic polar coordinates on polygonal
  meshes,'' \emph{Computer Graphics Forum}, vol.~31, no.~8, pp. 2423--2435,
  2012.

\bibitem{DBLP:journals/cgf/BelyaevF15}
A.~G. Belyaev and P.~Fayolle, ``On variational and pde-based distance function
  approximations,'' \emph{Comput. Graph. Forum}, vol.~34, no.~8, pp. 104--118,
  2015.

\bibitem{PSHM}
J.~Tao, J.~Zhang, B.~Deng, Z.~Fang, Y.~Peng, and Y.~He, ``Parallel and scalable
  heat methods for geodesic distance computation,'' \emph{IEEE Transactions on
  Pattern Analysis and Machine Intelligence}, vol.~43, no.~2, pp. 579--594,
  2021.

\bibitem{dgg}
X.~Wang, Z.~Fang, J.~Wu, S.~Xin, and Y.~He, ``Discrete geodesic graph {(DGG)}
  for computing geodesic distances on polyhedral surfaces,'' \emph{Comput.
  Aided Geom. Des.}, vol.~52, pp. 262--284, 2017.

\bibitem{XIN2018Lightweight}
S.~Xin, W.~Wang, Y.~He, Y.~Zhou, S.~Chen, C.~Tu, and Z.~Shu, ``Lightweight
  preprocessing and fast query of geodesic distance via proximity graph,''
  \emph{Computer-Aided Design}, vol. 102, pp. 128--138, 2018, proceeding of SPM
  2018 Symposium.

\bibitem{cox2000multidimensional}
T.~F. Cox and M.~Cox, \emph{Multidimensional Scaling, Second Edition},
  2nd~ed.\hskip 1em plus 0.5em minus 0.4em\relax Chapman and Hall/CRC, 2000.

\bibitem{DBLP:conf/smi/SorkineC04}
O.~Sorkine and D.~Cohen{-}Or, ``Least-squares meshes,'' in \emph{2004
  International Conference on Shape Modeling and Applications {(SMI} 2004), 7-9
  June 2004, Genova, Italy}.\hskip 1em plus 0.5em minus 0.4em\relax {IEEE}
  Computer Society, 2004, pp. 191--199.

\bibitem{Coifman7426}
R.~R. Coifman, S.~Lafon, A.~B. Lee, M.~Maggioni, B.~Nadler, F.~Warner, and
  S.~W. Zucker, ``Geometric diffusions as a tool for harmonic analysis and
  structure definition of data: Diffusion maps,'' \emph{Proceedings of the
  National Academy of Sciences}, vol. 102, no.~21, pp. 7426--7431, 2005.

\bibitem{commutedistance}
F.~{Fouss}, A.~{Pirotte}, J.~{Renders}, and M.~{Saerens}, ``Random-walk
  computation of similarities between nodes of a graph with application to
  collaborative recommendation,'' \emph{IEEE Transactions on Knowledge and Data
  Engineering}, vol.~19, no.~3, pp. 355--369, 2007.

\bibitem{LipmanRF10}
Y.~Lipman, R.~M. Rustamov, and T.~A. Funkhouser, ``Biharmonic distance,''
  \emph{{ACM} Trans. Graph.}, vol.~29, no.~3, pp. 27:1--27:11, 2010.

\bibitem{levy2013variational}
B.~L{\'e}vy and N.~Bonneel, ``Variational anisotropic surface meshing with
  voronoi parallel linear enumeration,'' in \emph{Proceedings of the 21st
  international meshing roundtable}.\hskip 1em plus 0.5em minus 0.4em\relax
  Springer, 2013, pp. 349--366.

\bibitem{2018Computing}
Z.~Zhong, W.~Wang, B.~Levy, J.~Hua, and X.~Guo, ``Computing a high-dimensional
  euclidean embedding from an arbitrary smooth riemannian metric,'' \emph{ACM
  Transactions on Graphics}, vol.~37, no. 4CD, pp. 1--16, 2018.

\bibitem{Bouaziz2012}
S.~Bouaziz, M.~Deuss, Y.~Schwartzburg, T.~Weise, and M.~Pauly, ``{Shape-Up}:
  {Shaping} discrete geometry with projections,'' \emph{Computer Graphics
  Forum}, vol.~31, no.~5, pp. 1657--1667, 2012.

\bibitem{LiuBK17}
T.~Liu, S.~Bouaziz, and L.~Kavan, ``Quasi-newton methods for real-time
  simulation of hyperelastic materials,'' \emph{{ACM} Trans. Graph.}, vol.~36,
  no.~3, pp. 23:1--23:16, 2017.

\bibitem{nocedal2006numerical}
J.~Nocedal and S.~J. Wright, \emph{Numerical optimization}, 2nd~ed.\hskip 1em
  plus 0.5em minus 0.4em\relax Springer-Verlag New York, 2006.

\bibitem{ceres-solver}
S.~Agarwal, K.~Mierle, and Others, ``Ceres solver,''
  \url{http://ceres-solver.org}.

\end{thebibliography}
